# DIBUJANDO SUPERFICIES EN PERSPECTIVA CON ELIMINACIÓN DE LÍNEAS OCULTAS

*En honor a mi profesor el Dr. Olgierd Alf Biberstein **

**PERSPECTIVE DRAWING OF SURFACES WITH LINE HIDDEN LINE ELIMINATION**

Ignacio Vega-Páez[1], José Angel Ortega[2] y Georgina G. Pulido[3]
ivega@g-ibp.com, oeha430210@hotmail.com y gpulido@att.net.mx

## ABSTRACT

An efficient computer algorithm is described for the perspective drawing of a wide class of surfaces. The class includes surfaces corresponding lo single-valued, continuous functions which are defined over rectangular domains. The algorithm automatically computes and eliminates "hidden lines." The number of computations in the algorithm grows linearly with the number of sample points on the surface to be drawn. An analysis of the algorithm is presented, and extensions lo certain multi-valued functions are indicated. The algorithm is implemented and tested on .Net 2.0 platform that left interactive use. Running times are found lo be exceedingly efficient for visualization, where interaction on-line and view-point control, enables effective and rapid examination of a surfaces from many perspectives.

## RESUMEN

Se describe un algoritmo eficiente para dibujar una clase de superficies en perspectiva. La clase incluye superficies que corresponden a funciones continuas uni-valuadas, definidas sobre dominios rectangulares. El algoritmo en automático calcula y elimina la "líneas

[1] Sección de Estudios de Posgrado e Investigación de la Escuela Superior de Ingeniería Mecánica y Eléctrica del Instituto Politécnico Nacional (SEPI ESIME IPN) y Dirección de Tecnologías de la Información de la Empresa International Business Partner Consulting (IBP -Consulting)

[2] Sección de Estudios de Posgrado e Investigación de la Escuela Superior de Ingeniería Mecánica y Eléctrica del Instituto Politécnico Nacional (SEPI ESIME IPN)

[3] Departamento de Ciencias Básicas e Ingeniería de la Universidad Autónoma Metropolitana (UAM).

* Profesor de la Escuela Superior de Física y Matemáticas del Instituto Politécnico Nacional, de 1970 hasta su muerte, ocurrida en 1997

ocultas." El número de operaciones en el algoritmo crece linealmente con el número de puntos (ó muestra de puntos) de la superficie a ser dibujada. Se presentan tanto el algoritmo, así como una extensión a ciertas funciones multi-valuadas. El algoritmo es implantado y probado en plataforma .Net 2.0 que permite su uso interactivo. Se encuentran tiempos de ejecución eficientes que permiten la visualización, donde la interacción y el control del punto de observación, proporcionan una rápida visualización de superficies desde muchas perspectivas.

## INTRODUCCIÓN

El problema general de la eficiencia y significativamente el desplegando de objetos tridimensionales en dos dimensiones es central en la graficación por computadora. En particular, al dibujar objetos en perspectiva el "problema de ocultamiento de líneas" se realiza eliminando segmentos de línea no visibles desde el punto de vista del observador, este es uno de los problemas que siempre ha estado presente. A través del tiempo, varios algoritmos que tratan este problema han aparecido en la literatura. Algunos de estos algoritmos son prohibitivos por la gran cantidad tiempo de calculo (con respecto al número de puntos) o tienen requisitos del memoria muy grandes; algunos tienen ambas características indeseables.

En este artículo se da una descripción detallada de un algoritmo muy eficaz para dibujar en perspectiva una superficie arbitraria que corresponde a una función continua uni-valuada definida sobre un dominio rectangular, con la eliminación de líneas ocultas. El punto de observación desde donde es vista la superficie que pude ser cualquier punto fuera de la superficie. Se mencionan algunas generalizaciones del algoritmo a dominios no rectangulares y a funciones multi-valuadas.

Este algoritmo se ha implantado sobre la plataforma .Net 2.0 con tiempo de cómputo considerablemente más corto que algoritmos anteriores para dibujar superficies similares. Un rasgo muy útil de la aplicación es que el programa funciona interactivamente y permite al usuario cambiar los puntos de observación con el "ratón" e inmediatamente desplegar la vista de la superficie en análisis.

## II. FORMULACION DEL PROBLEMA

Dada una superficie tridimensional *S* definida sobre un dominio rectangular *R*, y un punto de observación *V* fuera de *S* (Fig. l), podemos asumir que *R* es centrado en origen *O* del plano *x-y* y orientado de forma que sus lados sean paralelos al eje *x* y al eje *y*, respectivamente. (Esto puede lograrse sin pérdida de generalidad por convenientes translación y rotación, tanto de *R* como de *V*).

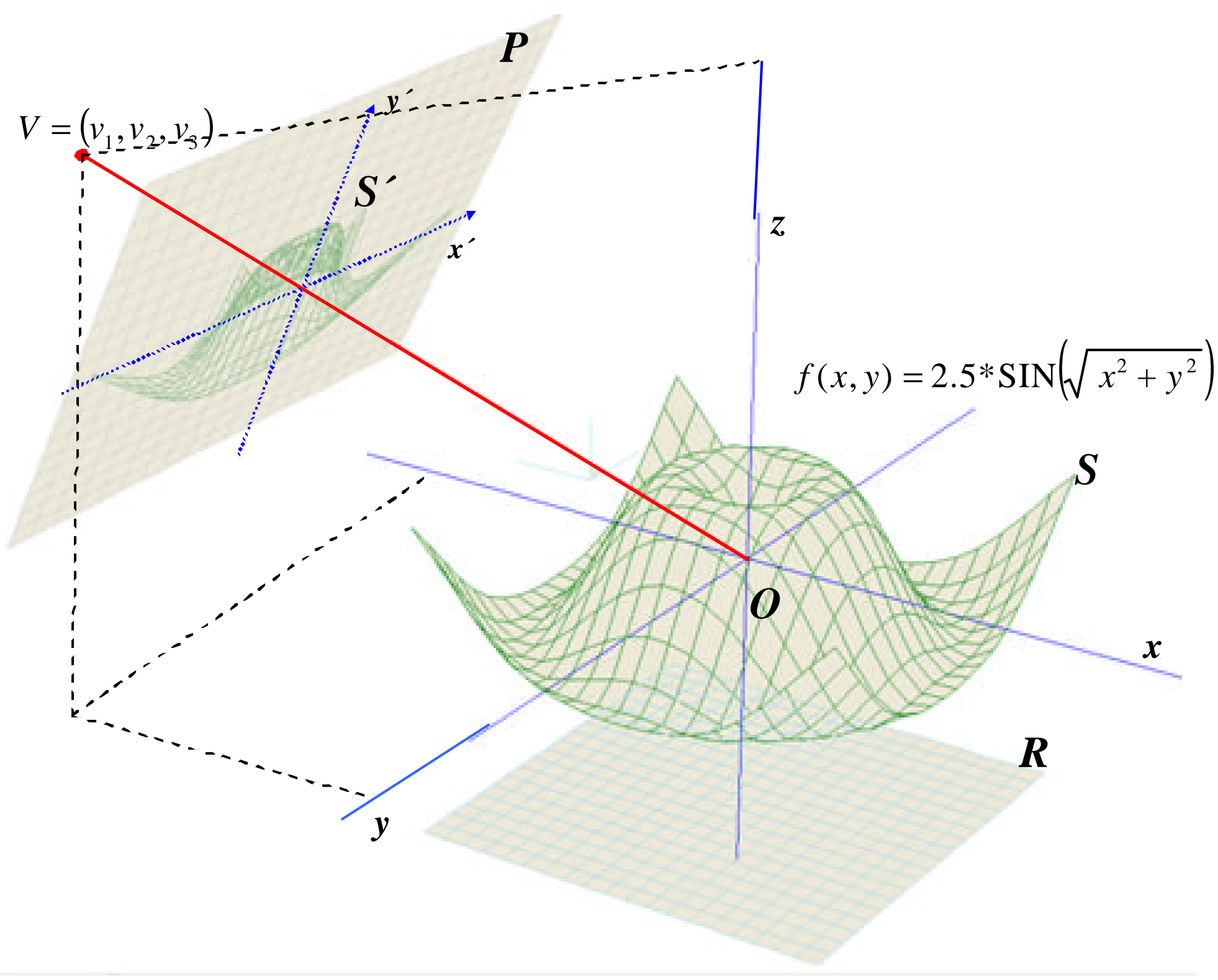

**Fig. 1. Proyección perspectiva de la superficie *S* en el plano *P*.**

El *P* plano, contiene la imagen en perspectiva *S'* de *S*, se elige perpendicular a la línea que une a *V* y el origen *O*. Se elige un sistema de coordenadas rectangular con ejes *x'*-*y'* en *P* que conserve la dirección vertical original de *S*. Es decir, el eje *y'* sobre *P* debe ser la proyección del eje *z* original. Esta última elección es necesaria, como lo veremos después. (Con relación al sistema rectangular sobre *P*, la imagen en perspectiva sobre *P* de cada punto de *S* tiene un par bien definido de coordenadas. La demostración de estas coordenadas se da en el Apéndice.)

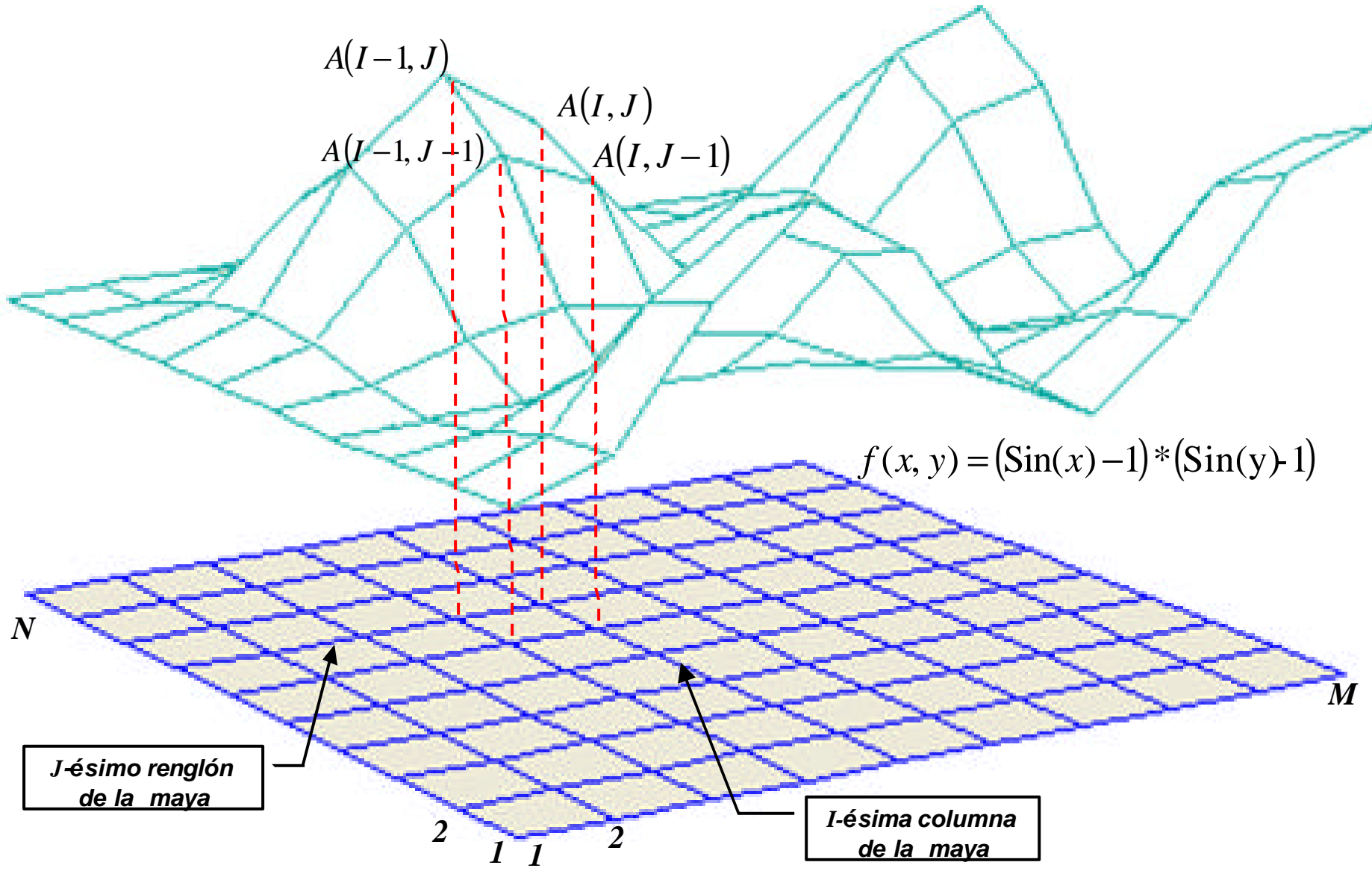

**Fig. 2. Parche de la superficie de un subrectángulo de la maya**

La superficie *S* es primero cuantificada en una maya de puntos rectangular *MxN* sobre el dominio *R*. Figura 2 muestras las líneas de la maya en *R*, dividiendo *R* en subrectangles. Se muestra la parte de *S* definida sobre un subrectangle. Será llamado como “el parche de la superficie” de *S*. Después de la cuantificación, sólo los cuatro puntos del parche definidos sobre los cuatro vértices del subrectangle son conocidos, y es asumida la interpolación lineal de la función entre puntos adyacentes de la maya, es decir, entre los puntos adyacentes en la misma fila y los puntos adyacentes en la misma columna de la maya. (Así, *S* no necesita ser definida explícitamente por una función matemática; los datos para *S* pueden ser dados por un arreglo rectangular de puntos que corresponden a una maya rectangular. de puntos sobre el dominio.) El comportamiento de *S* en el interior de cada subrectangle de la maya es ignorado. Así, cada parche de la superficie de *S* será representado por sus cuatro bordes linealmente interpolados en el espacio tridimensional. La imagen de estos bordes en el plano de proyección es un polígono de cuatro lados (Fig. 3). Para cada, parche de la superficie de *S*, sólo se dibujarán los segmentos de la línea visibles de estos bordes.

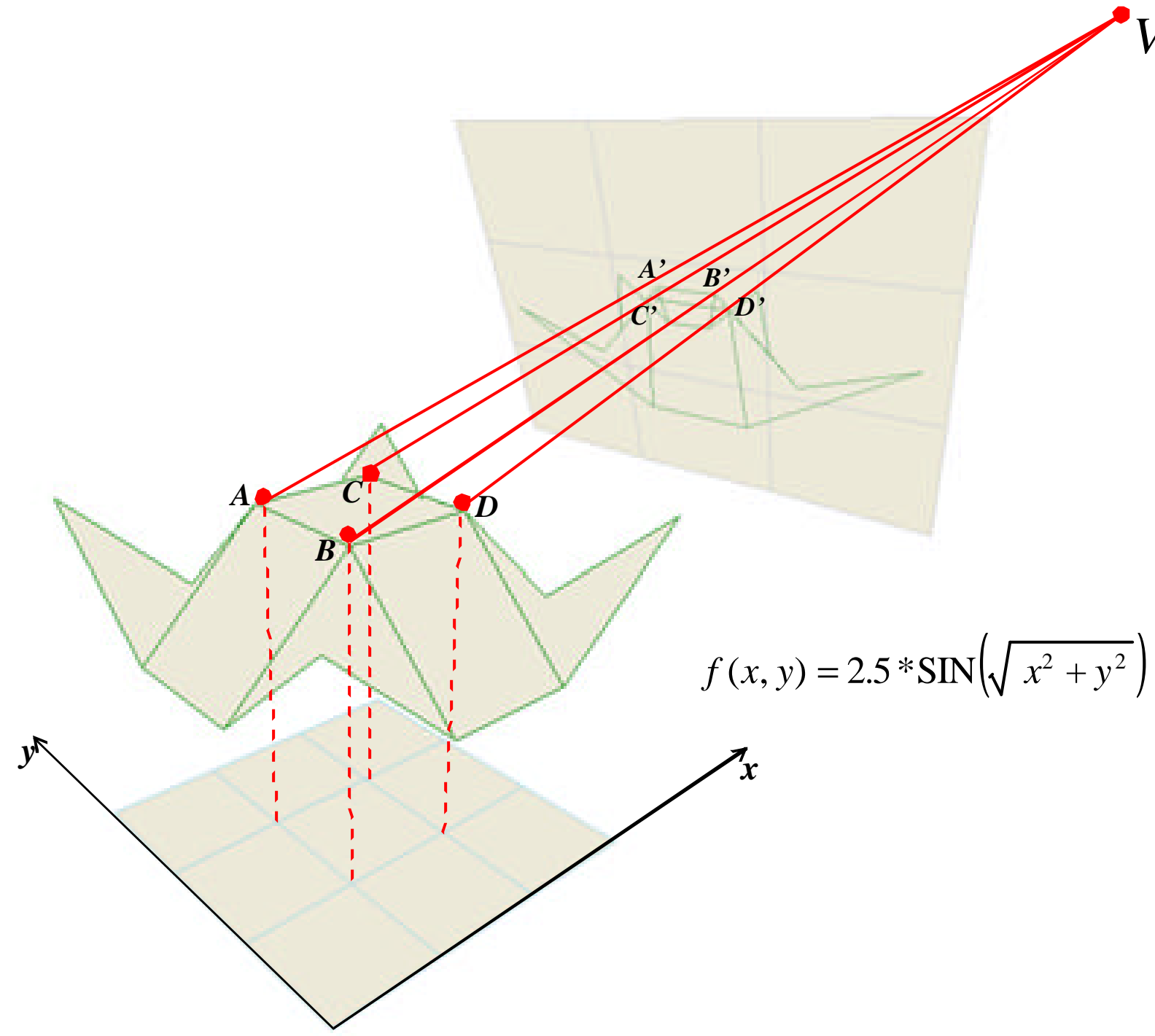

**Fig. 3. Proyección del parche de la superficie linealmente interpolado.**

Intuitivamente, la superficie debe pensarse de como una membrana elástica opaca estirada sobre un marco rígido que consiste en todos los bordes linealmente interpolados de los parches de la superficie. Entonces el problema es dibujar en perspectiva la superficie eliminando los segmentos de línea ocultos por la membrana opaca desde el punto de vista del observador. (Note que la proyección vertical sobre el plano *x-y* coincide con las líneas de la maya en *R*.)

## III. IDEAS BÁSICAS DEL ALGORITMO

El algoritmo para determinar la visibilidad de cualquier borde o segmento dado esta basado en dos ideas. La primera consiste en escoger un "ordenamiento" en particular de los parches de la superficie *S* para que ningún parche aparezca antes en el "orden" y que este oculto desde el punto de observación o que el parche aparezca después. La superficie debe ser dibujada, un parche a la vez acorde a ordenamiento elegido. Entonces, *un segmento de línea sobre algún borde de un parche bajo consideración es oculto desde el punto de observación, si y sólo si su imagen sobre el plano de proyección cae dentro de una región del plano ya cubierta por las imágenes de parches antes dibujados.*

Este particular ordenamiento de los parches de la superficie tienen la propiedad que, para una superficie $S$ que corresponde a una función continua univaluada sobre un dominio rectangular, la región del plano de proyección que va siendo cubierta por la imagen emergiendo de $S$ crecerá continuamente con cada parche que es dibujado en el orden elegido. Y debido a que se preserva la dirección vertical de la superficie, en cada fase del dibujo de $S$, la región del plano "oculta" puede limitarse con precisión entre dos funciones continuas lineales por partes. Por lo que la determinar de la visibilidad de cualquier borde de un parche de la superficie se reduce al problema de decidir qué parte de su imagen sobre el plano de proyección esta entre estas dos funciones. Esta delineación de la región oculta del plano de proyección, después de dibujar cada parche, por medio de las dos funciones continuas lineales por partes, constituye la segunda idea del algoritmo. Como se puede ver en la Fig 4.

Comentario: si el eje $y$ sobre el plano de proyección no se elige para ser la imagen perspectiva del eje $z$ original, entonces la imagen de la superficie será "inclinado" tal que será imposible de delinear su frontera usando sólo dos funciones.

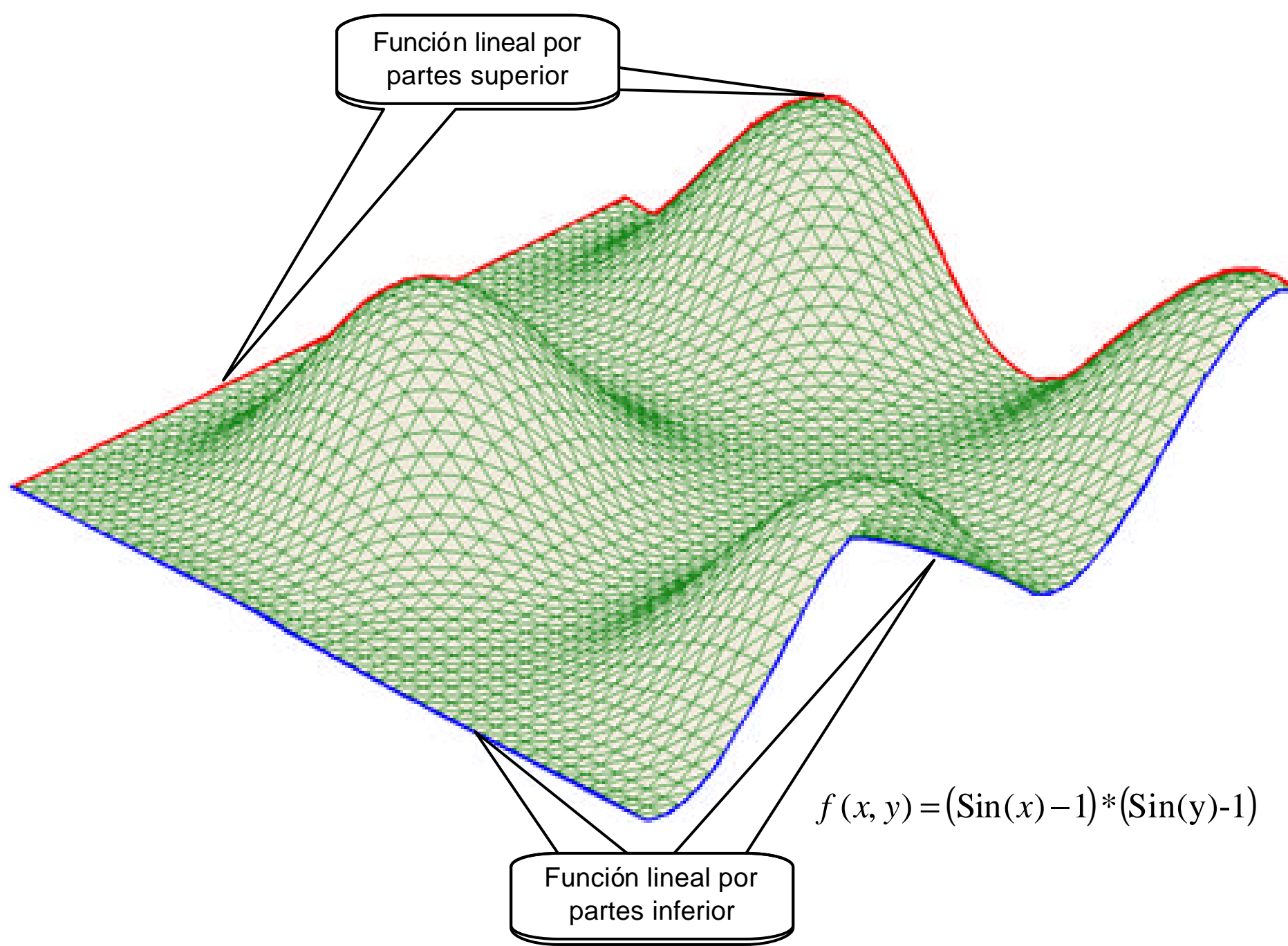

**Fig. 4. Región del plano "oculta" esta limitada entre dos funciones continuas lineales por partes.**

## IV. ORDENACION DE LOS PARCHES DE LA SUPERFICIE

Del hecho que es posible ordenar los parches de la superficie $S$ tal que los más cercanos no sean ocultados por los otros mas alejados depende de la siguiente observación:

Suponga $A$ y $B$ dos puntos arbitrarios de la superficie tal que $A$ oculta a $B$ desde $V$. Geométricamente, esto significa que $A$, $B$, y $V$ están en línea recta, y la distancia de $A$ a $V$ es menor que la distancia de $B$ a $V$. Entonces puede mostrarse fácilmente que las proyecciones verticales $A_0$, $B_0$, $V_0$, de los puntos $A$, $B$, $V$ respectivamente, sobre el plano $x$-$y$ satisfacen la también la relación. Es decir, son colineales, y $A_0$, esta más cercano que $B_0$, a $V_0$.

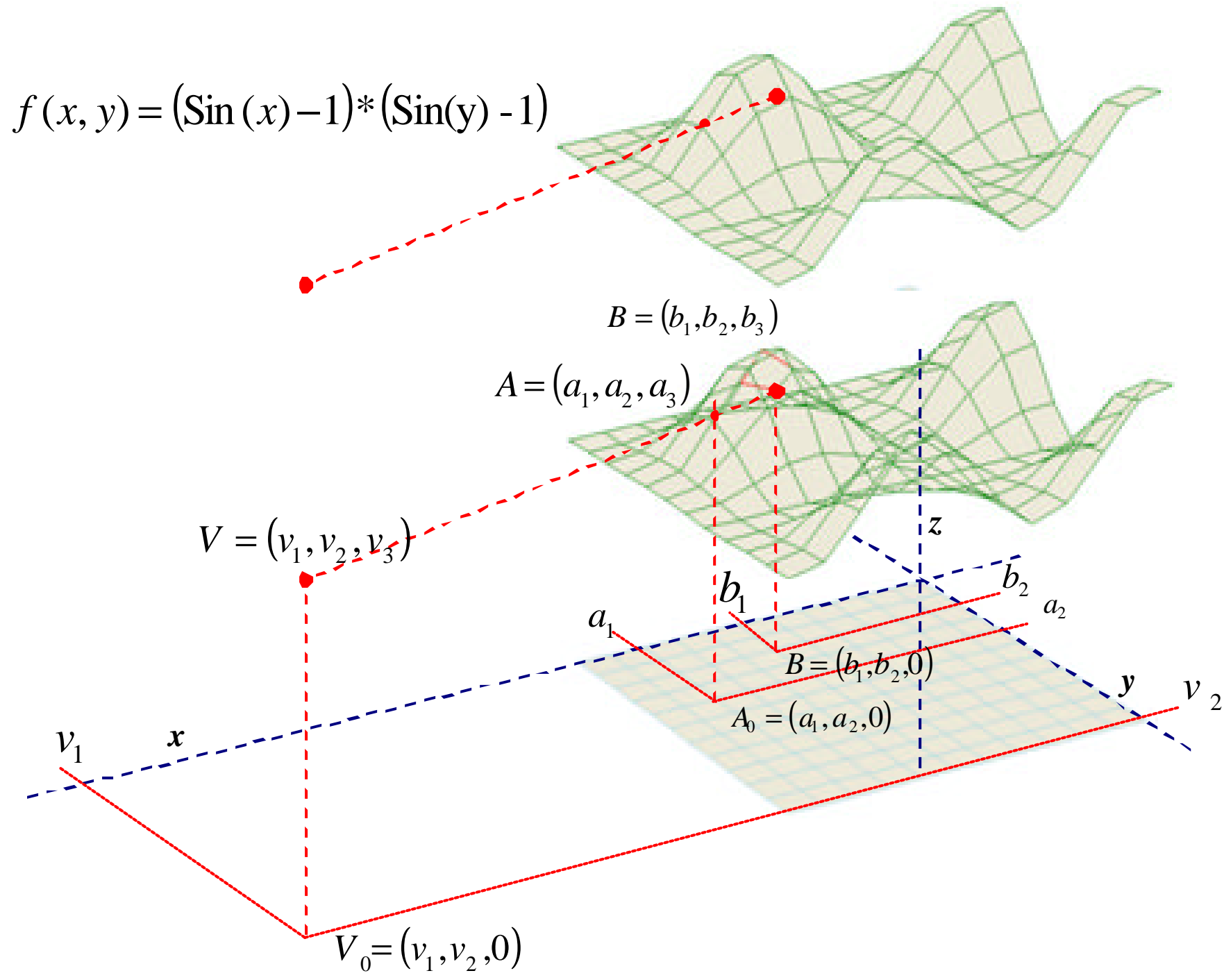

**Fig. 5 Los puntos *A* y *B* de la superficie con *A* ocultando a *B*.**

Esta observación es la clave para el ordenamiento de los parches de la superficie. Primero, considere el caso en que la proyección vertical $V_0=(v_1,v_2,0)$ del punto de observación hacia el plano $x$-$y$, es el suroeste del dominio, como se muestra en la figura Fig. 5. [Es decir, $v_1<=x_1$, y $v_2<=x_2$ , donde $(x_1,y_2,0)$ es la esquina izquierda más baja del dominio.] Una consecuencia inmediata de la observación anterior es que si $A=(a_1,a_2,a_3)$ oculta a $B=(b_1,b_2,b_3)$ desde $V=(v_1,v_2,v_3)$, entonces $a_1<b_1$, y $a_2<b_2$ ; i.e., $A_0$ debe estar más al suroeste de $B_0$ (consecuencia de da la localización suroeste de $V_0$). Esto sugiere la regla siguiente para determinar el orden relativo de puntos de $S$, dada la localizaron sudoeste de $V_0$: Sea $A$ y $B$ cualesquiera dos puntos de $S$. Si uno de ellos, digamos $A$, esta más al suroeste que el otro (más precisamente, si $A_0$, esta más al suroeste, de $B_0$), entonces $A$ debe preceder a $B$. Por otra parte, $A$ y $B$ pueden ordenarse independientemente de uno del otro.

Hay varias maneras de ordenar los parches de la superficie $S$ para que esta regla de orden relativo de puntos sea satisfecha, para que los primeros parches no sean ocultados por parches posteriores ver Figuras 6 y 7 que ilustran posibles ordenaciones de los subrectangulos de $R$. El ordenamiento inducido de los parches de $S$ tenga esta propiedad; i.e., ordenados por filas por filas Fig. 6, de izquierda a derecha en cada fila, empezando con la fila del frente (relativa a la localización suroeste de $V_0$). Es obvio que, bajo este ordenamiento de parches, ningún punto de un parche más cercano será ocultado por un punto de un parche posterior, desde que ningún punto de un parche cercano esta más al suroeste que cualquier punto de un parche posterior.

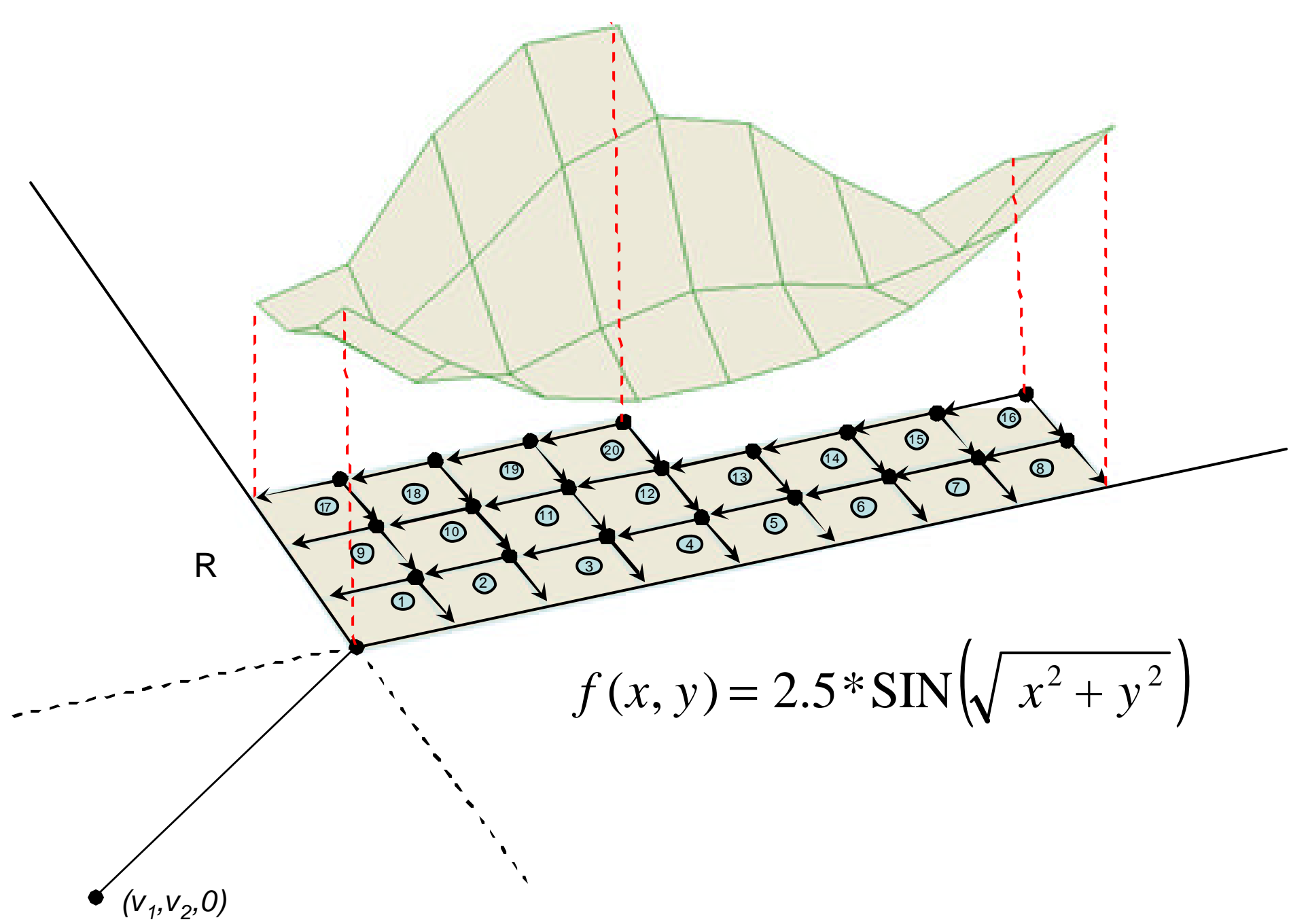

**Fig. 6 Ordenamiento frontal de los parches de la maya.**

Para otra "esquina" las situaciones de $V_0$, (i.e., noroeste, noeste, y sureste del dominio) pueden hacerse clasificaciones análogas de los parches de la superficie. En lugar de discutir clasificaciones convenientes de los parches de la superficie para otras situaciones de $V_0$, nosotros procederemos describir el algoritmo en detalle para el caso especial en que $V_0$, es suroeste del dominio, y entonces muestra cómo los otros casos pueden reducirse a casos que involucran sólo "esquina" las situaciones de $V_0$. Otro ordenamiento que preserva la propiedad se muestra en la Fig. 7 el ordenamiento de Cantor que recorre diagonalmente el dominio *R*, iniciando con la esquina mas cercana a $V_0$ o sea *(1,1), (1,N), (N,1) o (N,N)* lo cual garantiza que los mas cercanos se pintaran primero y los posteriores después.

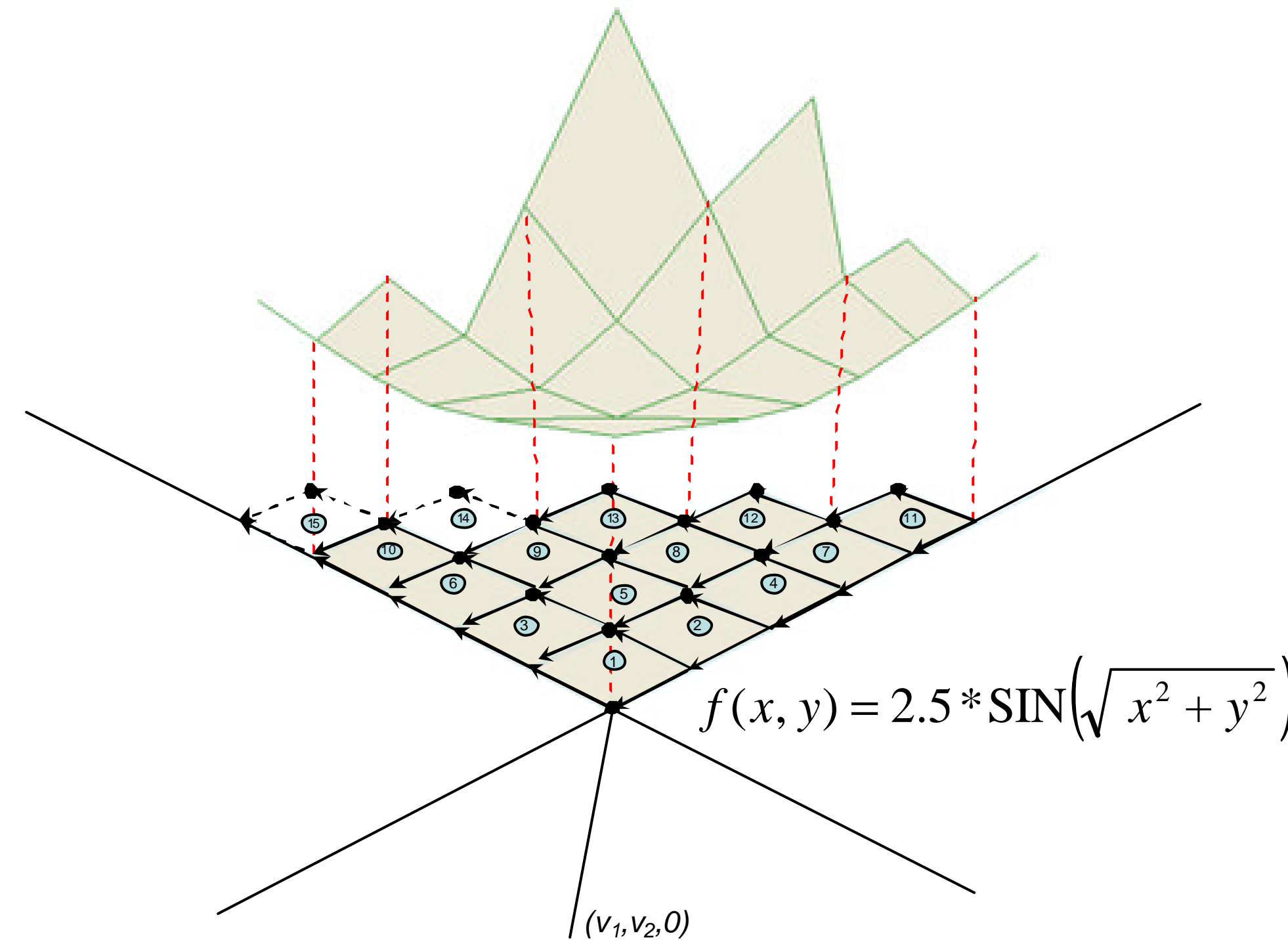

**Fig. 7 Ordenamiento de Cantor o diagonal de los parches de la maya.**

## V. EL ALGORITMO EN DETALLE

Sea *S* una superficie definida sobre un dominio rectangular *R* del plano *x-y*. Asumimos que *S* será visto en perspectiva desde un punto de observación *V* cuya la proyección $V_0$ vertical sobre el plano *x-y* es suroeste de *R*. Demos una maya *MxN* rectangular de puntos sobre en *R*, para *I=1,2,…,M* y *J=1,2,…,N*, sea *A(I, J)* que denote un "punto de la muestra" o simplemente punto sobre *S* definido sobre la maya *R* localizado en la columna *I-ésima* y renglón *J-ésimo* de la maya. Para *I=2,…,M* y *J=2,…,N,* sea *S(I, J)* el parche de la superficie con vértices *A(I, J), A(I,J-1), A(I-1,J), A(I-1,J-1).* (Ver Fig. 2.) Por encima de las consideraciones, los parches pueden ser ordenados como sigue, y dibujados según esta ordenación, uno por uno:

*S(2,2),S(3,2),…,S(M,2),*

$$S(2,3), S(3,3), \ldots, S(M,3),$$

.....

$$S(2,N), S(3,N), \ldots, S(M,N).$$

Ahora, considere la sucesión de puntos $A(1,J)$, $J=1,\ldots,N$, definidos a lo largo de la primera columna de puntos de la maya. El segmento de línea, que une estos puntos en orden consecutivo se llamara el “borde izquierdo principal” de $S$. Similarmente, el segmento de línea que une la sucesión de puntos $A(I,1)$, $I=1,\ldots,M$ en orden consecutivo se llamara el “borde frontal o derecho principal" de $S$. Como consecuencia de la observación hecha al principio de Sección IV, ambos bordes son completamente visibles desde $V$. Así, podemos empezar a dibujar todo el segmento de línea de los bordes principales. Entonces el dibujo de cada parche $S(I,J)$ siguiendo el orden mencionado puede ser agregando solo los segmentos visibles de dos de sus cuatro bordes, a saber su “borde trasero,” uniendo $A(I,J)$ y $A(I-1,J)$, y su "borde derecho," uniendo $A(I,J)$ y $A(I,J-1)$, porque los otros dos bordes ya debieron haber sido considerados en relación con los parches anteriores o como uno de los bordes principales.

A manera de ejemplo la Figura 8a da un ejemplo de una superficie definida sobre una maya de puntos *5x3*. El arreglo *5x3* de puntos muestra sobre la superficie da lugar a dos filas de parches, cuatro en la fila delantera y cuatro en la fila trasera. Figure 8b muestra la conjunción del borde izquierdo principal y el borde delantero principal de la superficie.

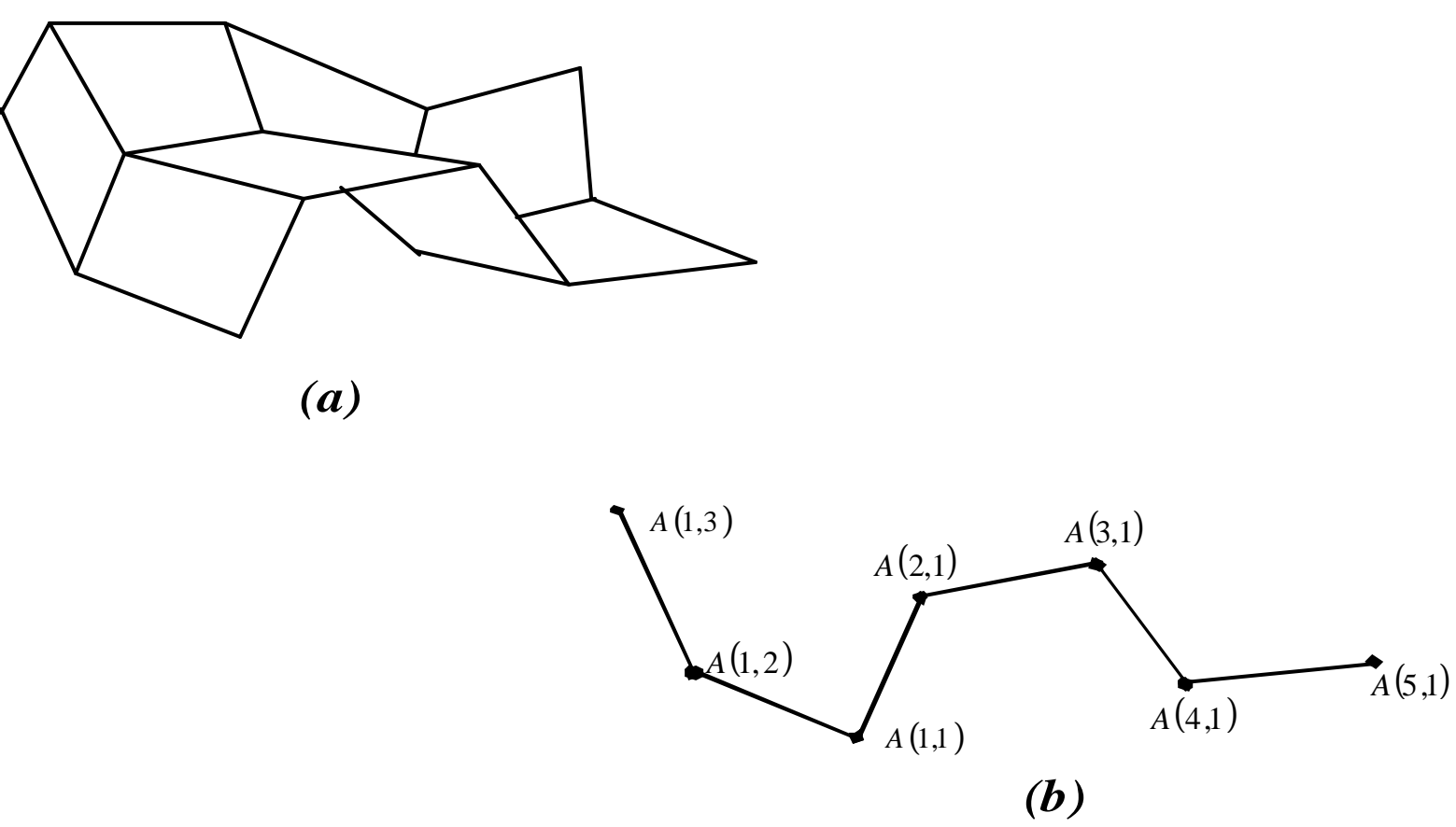

**Fig. 8, (a) superficie a manera de ejemplo y (b) bordes izq. y frontal principales (b)**

Figuras 9a a través de 9h muestra los parches siguientes que se agregan al dibujar la superficie. Con cada parche que se va agregándose, $L_1$, denota su borde “trasero”, y $L_2$. denota a su borde “derecho”. Las líneas punteadas indican los segmentos ocultos que no son pintados.

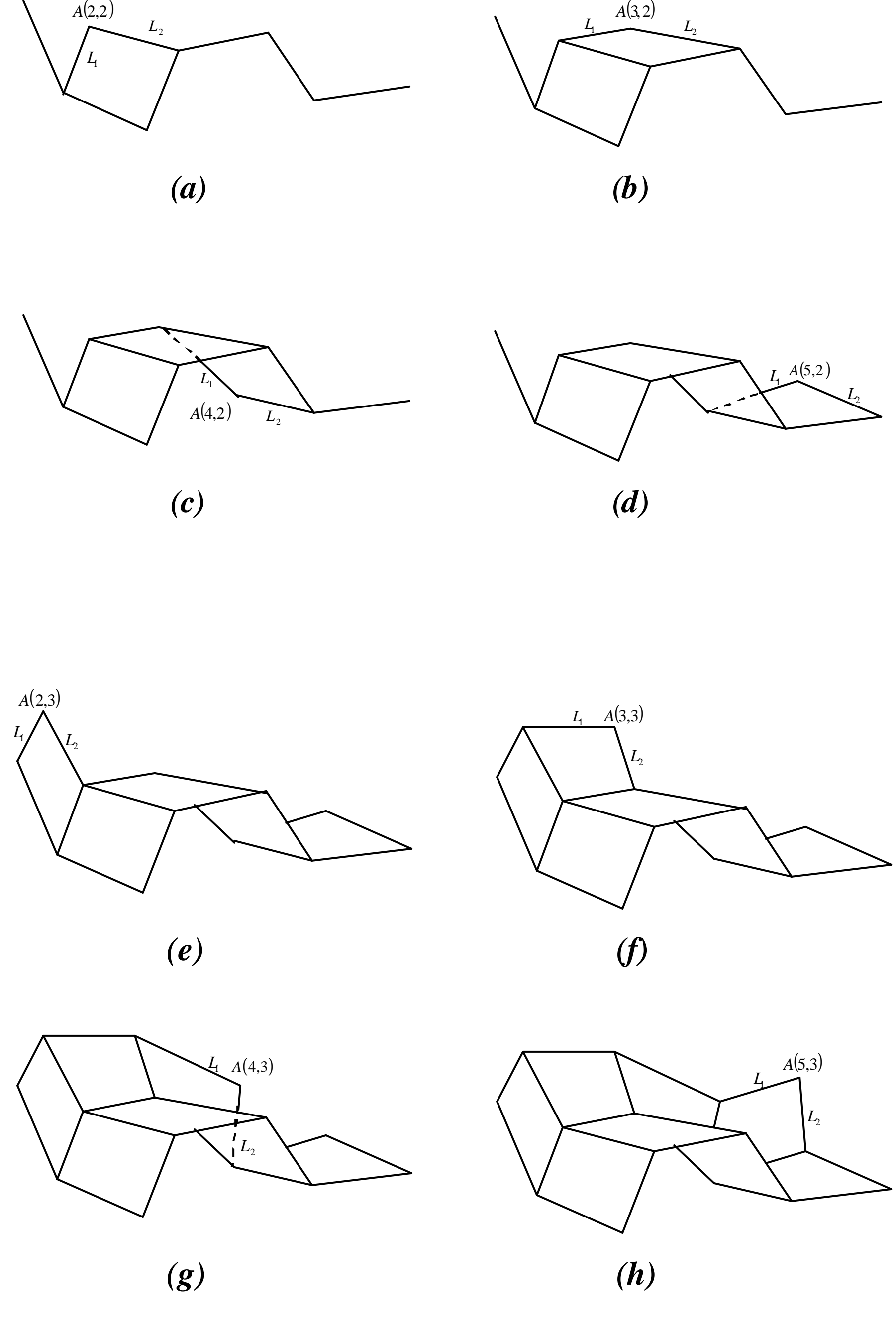

**Fig. 9, Secuencia en que sin pintados los parches de la superficie *S*.**

Como indicó con anterioridad, la determinación de la visibilidad de cada nuevo borde bajo consideración involucrará dos funciones continuas lineales por partes. Intuitivamente, se usará uno de estas funciones, *Max*, para definir la curva del perímetro superior de la región del plano de proyección hasta el momento cubierto, y la otra función, *Min*, se usará para definir la curva quebrada del perímetro más bajo. Hasta este punto, sería útil identificar la maya del dispositivo de trazando con un rectángulo en el plano de proyección que circunscribe a la imagen en perspectiva de la superficie. (Esto corresponde a enmarcar y transformar las coordenadas de los puntos de la imagen para que caigan dentro de las dimensiones de la maya del dispositivo de trazando.) Entonces para perfeccionar la precisión en el delineando la región oculta del plano de proyección, o equivalentemente, la región oculta de la maya de trazando, el número de puntos de cada uno de las funciones *Max* y *Min* se escoge que coincida con el número de unidades horizontales de la maya del dispositivo de trazo.

Antes de iniciar el trazado, artificialmente ponemos $Max(k)=Min(k)=-1$, para $k=1,\ldots,K_x$, donde $K_x$ es la dimensión horizontal de la maya de trazando. Entonces, después de que cada segmento de la línea es dibujado, *Max* o *Min* (posiblemente ambos) serán modificados para reflejar el cambio en la forma de la imagen surgiendo de *S*. empezamos con los segmentos de la línea en los bordes principales *S*. Sea *A* y *B* cualesquiera dos "puntos muestra" consecutivos del borde izquierdo principal o del borde frontal principal, eso es, cualquier dos punto consecutivo en la sucesión siguiente (vea Fig. 2):

$$A(1,N), A(1,N-1), \ldots, A(1,1),\ A(2,1), \ldots,\ A(M,1)$$

Sea $A_x$, $A_y$, las coordenadas de trazando de *A* (escalados y trasladados), y $B_x$ $B_y$, Las coordenadas de trazando de *B*. Primero, el segmento de la línea que une los puntos $(A_x, A_y)$ y $(B_x, B_y)$ es pintado. Entonces se redefinen *Max* y *Min* entre $A_x$ y $B_x$ poniendo

$$Max(A_x)=Min(A_x)=A_y$$

$$Max(B_x)=Min(B_x)=B_y$$

e interpolando linealmente *Max* y *Min* linealmente $A_x$ and $B_x$. (En la Fig. 10, $K_x$. and $K_y$, refiere a las dimensiones de la maya de trazando, en las direcciones horizontales y verticales, respectivamente.)

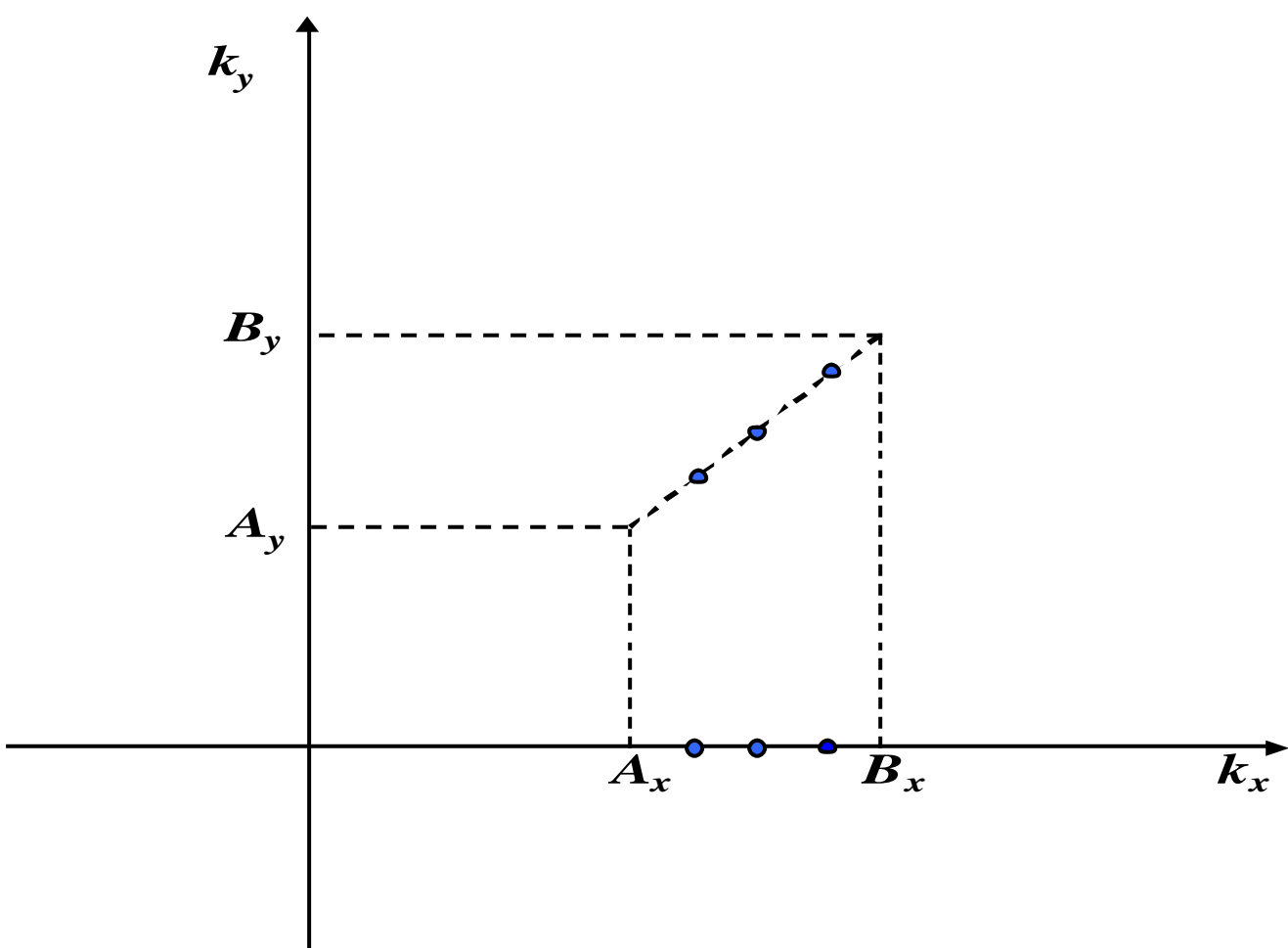

**Fig. 10, Interpolación lineal de Max y Min**.

Después de que el procedimiento anterior se lleva a cabo para cada par de puntos consecutivos en el ordenamiento, los bordes principales se dibujarán completamente, y su línea quebrada será definido por *Max* y *Min* entre los dos puntos finales. Note que la región limitada entre *Max* y *Min* tiene área cero y refleja el hecho que ningún parche ha sido todavía dibujado.

Ahora procedemos a dibujar los parches de la superficie. Cada *S(I,J)* en el ordenamiento, empezando con *S(2,2)* se procesará uniformemente como sigue: Permita $L_1$ denotar el borde entre *A(I,J)* y *A(I-1,J)*, $L_2$, denote el borde entre *A(I,J)* y *A(I,J-1)*. Para cada uno de estos bordes, debemos primero, determine sus segmentos visibles, entonces dibuje sólo esos segmentos, y modifique *Max* o *Min* para reflejar el cambio en la forma de la región oculta (el borde entero puede ocultarse, por supuesto).

Primero, Sea $L=L_1$, y *A=A(I,J)*, *B=A(I-1,J)*, los puntos finales de $L_1$. El criterio siguiente se usa para determinar la visibilidad de cualquier punto *C* sobre *L*: Si $C_x$ , $C_y$ son las coordenadas de trazo de *C*, entonces *C* es visible si y sólo si $C_y=Max(C_x)$ o $C_y=Min(C_x)$. Este criterio de visibilidad se aplica a cada uno de de los puntos finales *A* y *B*. (Sea $A_x$, $A_y$ las coordenadas de trazo de *A* y $B_x$, $B_y$ los de *B*.). Hay cuatro posibilidades:

(i) *A, B* son ambos visibles. Entonces asumimos que todos puntos de *L* son visibles. (Esta suposición se hace para reducir tiempo del cómputo. Puede pasar que se dibujen incorrectamente algunos segmentos de líneas ocultas si la superficie tiene afilados “púas” en primer plano. Este problema se le puede dar la vuelta iniciando con una maya más fina sobre *R*.). El segmento de la línea que une $(A_x,A_y)$ y $(B_x,B_y)$ es dibujado. Si $A_y=Max(A_x)$, *Max* se interpola linealmente entre $A_x$. y $B_x$ poniendo $Max(A_x)=A_y$, $Max(B_x)=B_y$ . De otra manera, debemos tener $A_y=Min(A_x)$, y *Min* se modifica similarmente entre $A_x$ y $B_y$, con $Min(A_x)=A_y$, $Min(B_x)=B_y$.

(ii) *A* y *B* son ambos ocultos. Entonces por las mismas razones como en el caso (i), asumimos que todos los puntos de *L* están ocultos y ningún segmento de la línea es dibujado arrastrado. *Max* y *Min* quedan inalterados, y procedemos con el siguiente borde.

(iii) *A* está oculto y *B* es visible. Una búsqueda se hace a lo largo de la línea que une *A* y *B*, empezando en *A*, para el primer punto visible *C* (se toman pasos discretos en la dirección horizontal de la maya de trazando). Se asume que el segmento que une a *C* con *B* es completamente visible y es dibujado. Si $C_y=Max(C_x)$, *Max* es modificado linealmente entre $C_x$ y $B_x$, con $Max(C_x)=C_y$, $Max(B_x)=B_y$, De otra forma, que *Min* se modifica entre $C_x$ y $B_x$.

(iv) *A* es visible y *B* está oculto. Una búsqueda se hace a lo largo de la línea que une *A* y *B* para el primer punto visible *C*, empezando del punto oculto de *B*. El resto es análogo al caso (iii).

Para cada parche de la superficie *S(I,J)*, este procedimiento se lleva a cabo con $L=L_1$ y $B=A(I\text{-}1,J)$; entonces el procedimiento se repite con $L=L_2$ y $B=A(I,J\text{-}1)$. Esto completa la descripción del algoritmo para el caso especial en el que la proyección vertical de $V_0$ del punto de observación sobre el plano *x-y*, esta en el suroeste del dominio *R*.

Si $V_0$ están en cualquiera de la otra "esquina" de las regiones del plano *x-y* (indicado por NW, NE, y SE en Fig. 11a), podemos rotar los datos que definen la superficie (o la función matemática o la serie rectangular de puntos muestra) por 90, 180, o 270 grados, respectivamente, y aplica el algoritmo como describió para $V_0$ para la región de SW del plano *x-y*. Otro acercamiento es cambiar el ordenamiento de los parches de la superficie y definir nuevos bordes principales para satisfacer las otras localizaciones de $V_0$.

| | | |
|---|---|---|
| ***NO*** | | ***NE*** |
| | ***R*** | |
| ***SO*** | | ***SE*** |

*(a)*

| | | |
|---|---|---|
| | ***N*** | |
| ***O*** | ***R*** | ***E*** |
| | ***S*** | |

*(b)*

**Fig. 11. Posibles localizaciones del punto de observación.**

Si $V_0$ está en cualquiera de las regiones directamente el oeste, norte, el este, o al sur de $R$ (Fig. 11b), podemos dividir $R$ en dos sub-rectángulos $R_1$ y $R_2$ (Fig. 12b), y aplica el algoritmo a cada uno de $R_1$, $R_2$, desde que $V_0$ es una "esquina" la localización de cada sub-rectángulo. Las dos partes de la superficie corresponden a $R_1$ y $R_2$, tendrá que ser "pegado a partes" a lo largo de la línea de partición, y esto puede hacerse sin "fallas" aumentando la muestra de puntos de la superficie con el conjunto de puntos definidos en la intersección de la línea de la partición con las líneas de la maya sobre $R$.

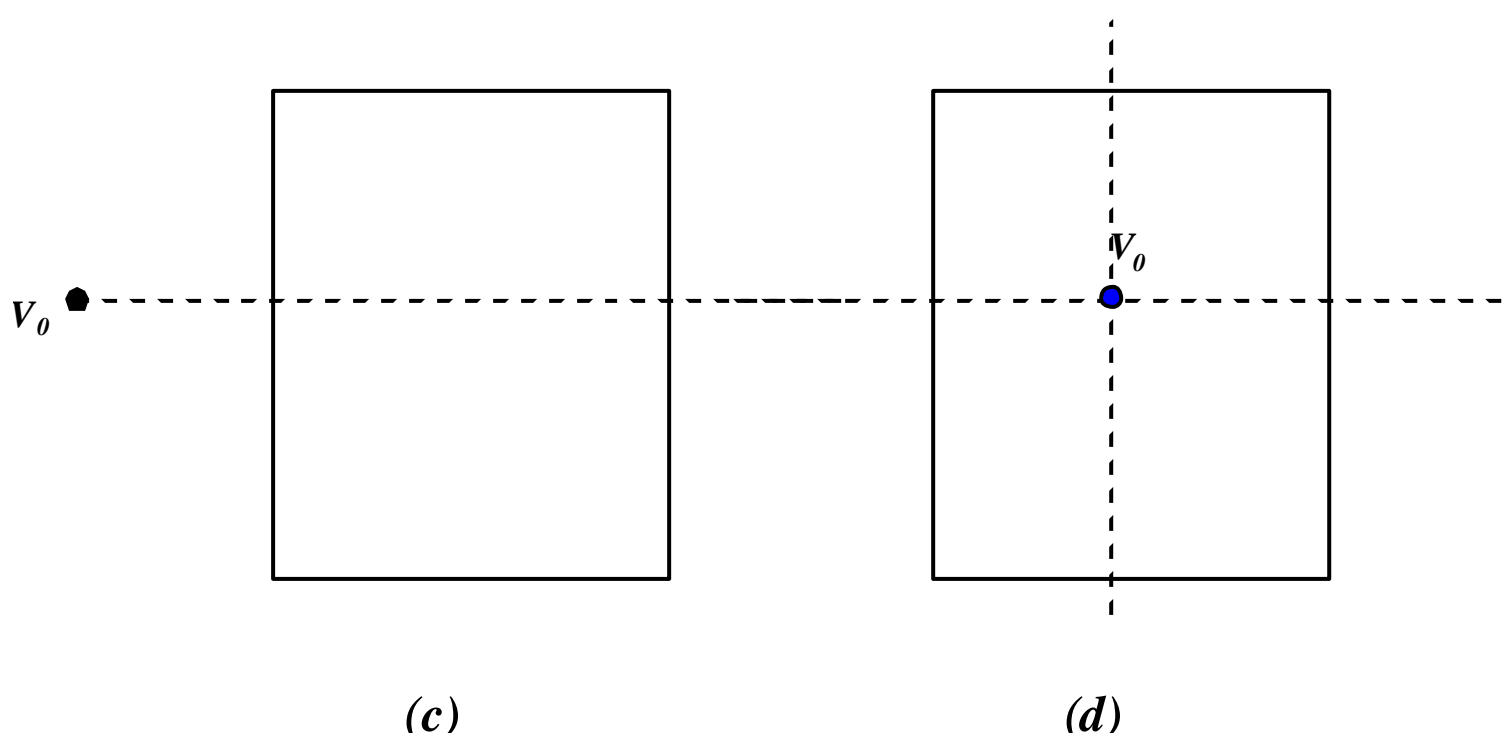

**Fig. 12. Particionado del dominio para localizaciones no esquinadas de $V_0$.**

Finalmente, si $V_0$ esta dentro del dominio *R*, podemos dividir a *R* en cuatro sub-rectángulos (Fig. 11d) y aplicar el algoritmo a cada uno de $R_1$, $R_2$, $R_3$, $R_4$, con $V_0$ situado en la "esquina".

Comentario: Note que bajo la suposición que el plano de proyección *P* debe se perpendicular a la línea que une a *V* y el origen, si $V_0$ coinciden con el origen, entonces la proyección del *z*-eje de *V* hacia *P* es un punto. En este caso, el *y'*-eje en *P* debe escogerse que sea la imagen de alguna otra línea. Vea el Apéndice.

Figuras 1 a 7 fueron generados por la implementación en computadoras PC compatibles con Net Framework. Y las superficies de la figuras 13 y 14 tambien muestran el ambiente de trabajo de la implementación, para estos cosos se evaluaron con una maya cuadrada de puntos 150 y 90, para un total de 22500 y 180 puntos.

Está claro que usando este algoritmo, el tiempo de cómputo debe variar linealmente con el número de puntos, esto es porque cada punto es procesado evaluando sus coordenadas de trazo *IX*, *IY*, y entonces la ordenada *IY* es comparada con dos valores, *Max(IX)* y *Min(IX)*.

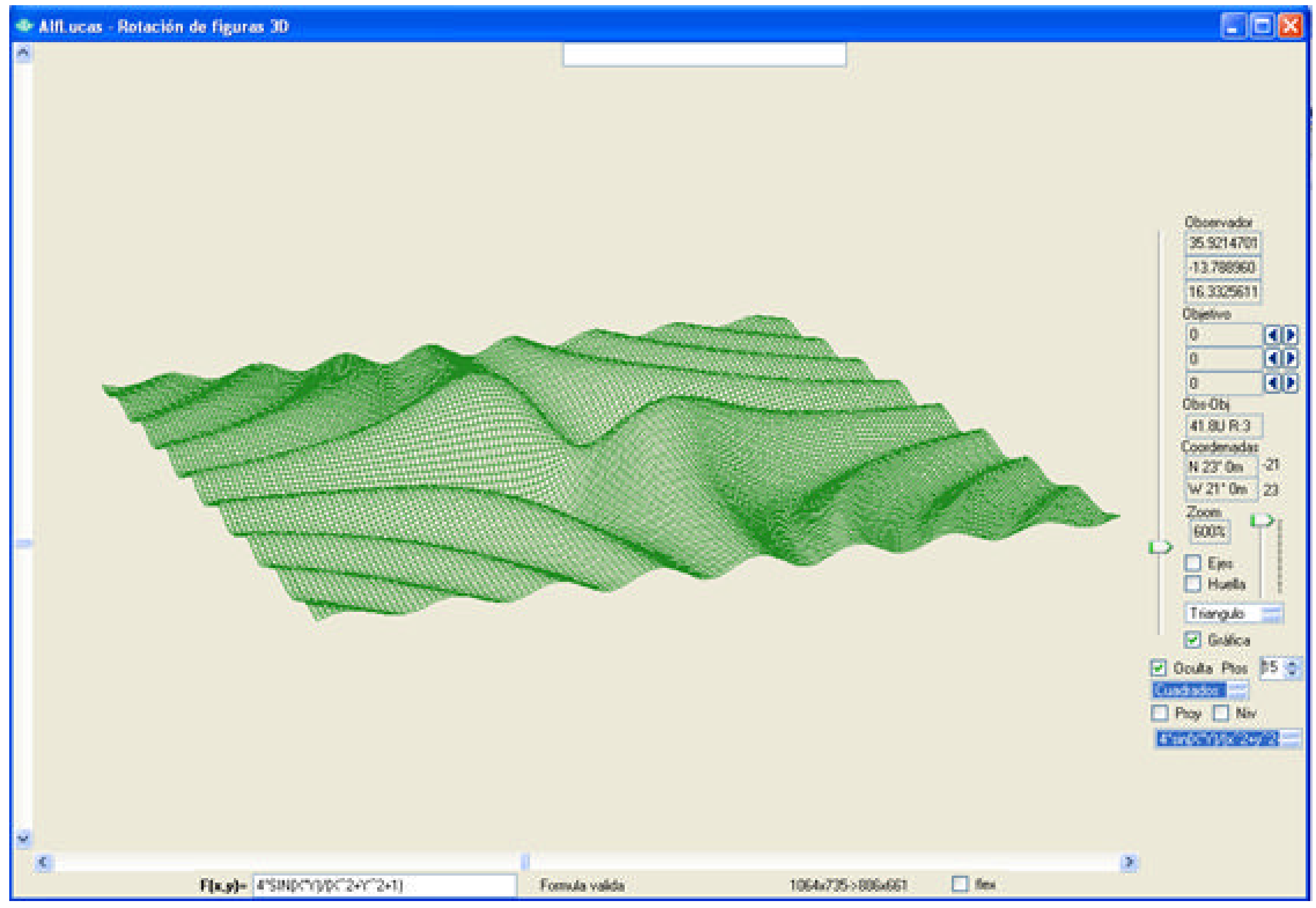

**Fig. 13 Ambiente de trabajo de la implementación..**

# VI. GENERALIZACIONES

## 6.1 Dominios convexos

El algoritmo puede generalizarse a una superficie arbitraria que $S$ definida por una función continua uni-valuada sobre un dominio convexo $R$, extendiendo la definición de la función $f$ artificialmente a un dominio rectangular $R'$ conteniendo $R$. Entones se aplican las ideas del algoritmo a la superficie $S'$ definido por la función extendida $f'$ encima de $R'$, con las modificaciones siguientes:

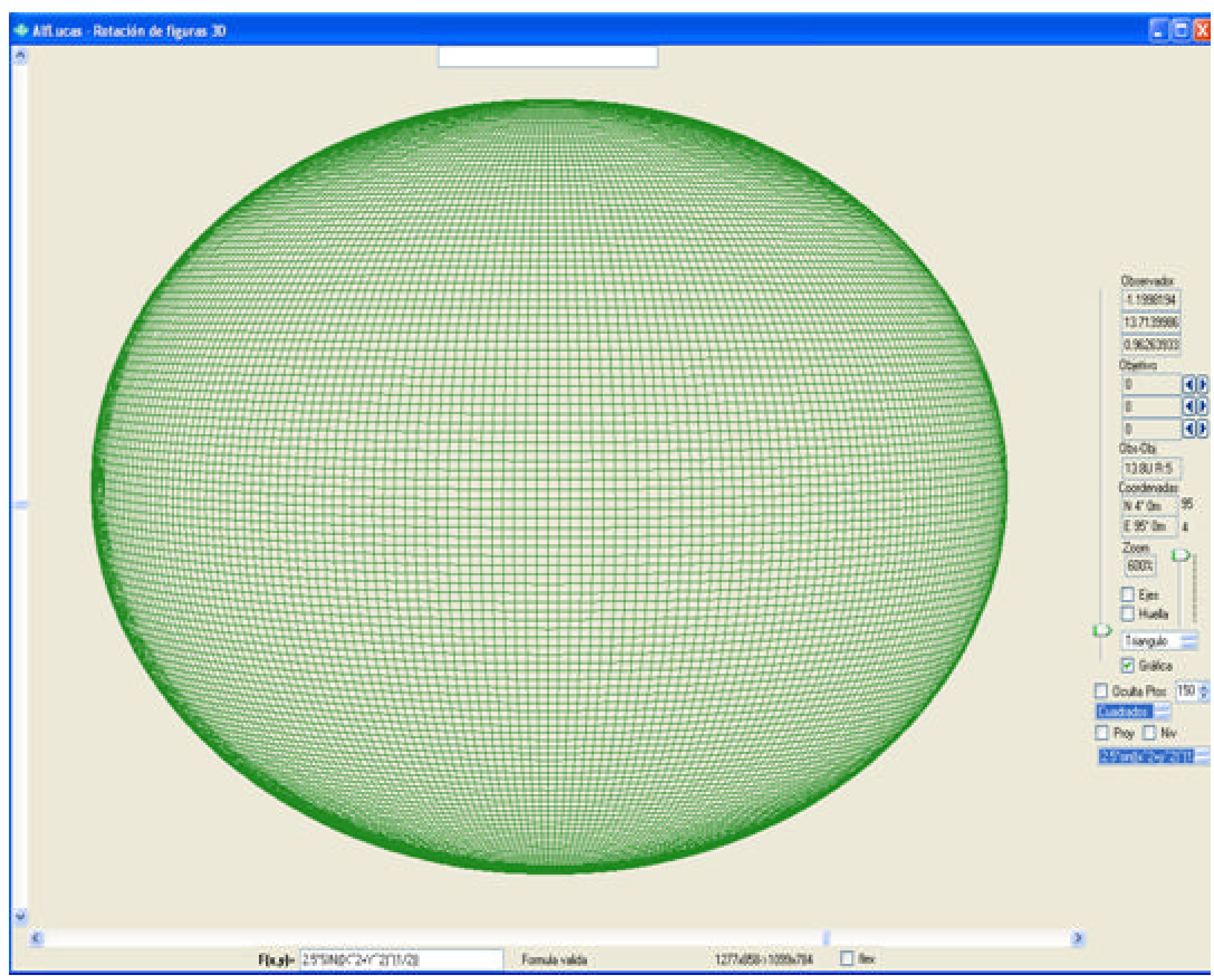

**Fig. 14 La esfera en el ambiente de implementación con dominio convexo.**

Primero, la muestra de puntos a lo largo de de los bordes principales de $S'$ debe ser reemplazado por un subconjunto de puntos definido en la intersección de la frontera de $R$ con las líneas de la maya sobre $R'$, y debe darse consideración especial al dibujar las líneas que conectan estos puntos.

Segundo, una prueba debe hacerse para la muestra de puntos de $S'$ para determinar si ellos pertenecen a $S$, para que sólo segmentos de la línea que son parte de $S$ sean pindados. Una manera de lograr esto está poniendo $f(x,y)=z_0$ para todo $(x,y)$ en $R'$-$R$, donde $z_0$ es un número fuera del rango de valores de $f$ sobre $R$. (La convexidad del dominio es necesaria para acotar la imágen que emerge en cada fase del dibujo entre dos funciones.)

6.2 Funciones multi-valuadas

El algoritmo también puede extenderse a ciertas funciones multi-valuadas. Por ejemplo, si la superficie es una esfera, esta es primero dividida por un plano horizontal en dos piezas uni-valuadas, y si el punto de observación es anterior al plano horizontal, el algoritmo es primero aplicado al hemisferio superior, y entonces, sin el re-inicializar *Max* o *Min*, el algoritmo se aplica al hemisferio inferior.

## APÉNDICE

Evaluación de Coordenadas Rectangulares de Puntos de la Imagen

Sea $S$ una superficie definida sobre un dominio centrado al origen del plano $x$-$y$, y sea $V=(v_1 , v_2 , v_3)$ un punto que no pertenenezca a $S$. Para simplificar el cálculo de las coordenadas de los puntos de la imagen y de reducir tiempo del cómputo, un plano especial $P^*$, normal a $\overline{V}$ y atravesando el origen, se usará como plano de proyección. (En Fig. 1, un $P$ plano paralelo a $P^*$ es mostrado. Note que la imagen de $S$ sobre $P$ difiere de la imagen de $S$ en $P^*$ sólo por un factor de escala.)

Sea $A=(a_1, a_2, a_3)$ cualquier punto en $S$, y sea $B=(b_1,b_2,b_3)$ que denote este punto imagen sobre $P^*$. Las coordenadas de trazo de $A$ dependen de las coordenadas bidimensionales $(u,v)$ de $B$ con respecto a un sistema del ortogonal en $P^*$. Si $\overline{X} = \overline{(x_1, x_2, x_3)}$ y $\overline{Y} = \overline{(y_1, y_2, y_3)}$ son los vectores unitarios positivos del sistema de coordenadas en $P^*$, entonces los escalares, $u$ y $v$ deben satisfacer la relación siguiente:

$$u\overline{X} + v\overline{Y} = \overline{B} \tag{1}$$

Desde que $B$ es co lineal con los puntos $A$ y $V$, entonces,

$$\overline{B} = \overline{V} + r\left(\overline{A} - \overline{V}\right), \text{ para algún escalar } r. \tag{2}$$

Así, ecuación . (1) puede reemplazarse por

$$u\overline{X} + v\overline{Y} = \overline{V} + r\left(\overline{A} - \overline{V}\right) \tag{3}$$

Tomando el producto del punto de ambos lados de ecuación (3) con $\overline{X}$ , tenemos

$$u\overline{X}\cdot\overline{X} + v\overline{Y}\cdot\overline{X} = \overline{V}\cdot\overline{X} + r\overline{A}\cdot\overline{X} - r\overline{V}\cdot\overline{X} \tag{4}$$

Se tiene $\overline{X}\cdot\overline{X} = 1$ y $\overline{Y}\cdot\overline{X} = \overline{V}\cdot\overline{X} = 0$ ($\overline{V}\cdot\overline{X} = 0$ porque $\overline{X}$ sobre *P** y *P** es normal a $\overline{V}$ ), tenemos

$$u = r\overline{A}\cdot\overline{X} \tag{4}$$

Similarmente, tomando el producto del punto de ambos lados de eq. (3) con $\overline{Y}$ , tenemos

$$v = r\overline{A}\cdot\overline{Y} \tag{5}$$

El escalar *r* esta determinado por tomar el producto del punto de ambos lados de ecuación. (2) con $\overline{V}$ . Se tiene que $\overline{B}$ esta sobre *P**, entonces $\overline{B}\cdot\overline{V} = 0$ , y como consecuencia tenemos

$$r = \overline{V}\cdot\overline{V} / \overline{V}\cdot\left(\overline{V} - \overline{A}\right) \tag{6}$$

Falta por determinar las coordenadas de los vectores unitarios $\overline{X}$ , $\overline{Y}$ .

Como se mencionó en la Sección II, el eje *y* sobre *P** debe ser la imagen perspectiva del eje *z* original. Note, sin embargo, que si *V* está sobre el eje *z* entonces la imagen perspectiva de *V* del eje *z* es un solo punto, que por supuesto no puede usarse como línea eje del eje. En este caso, debemos componer esto moviendo el punto de observación ligeramente del eje *z* o escogiendo el eje *y* sobre *P** para ser la imagen de un ligero “oblicuo” del eje *z*. Para el resto de esta discusión asumimos que *V* no está en el eje *z.*

Permita *Z=(0,0,1).* Entonces, $\overline{V}$ , $\overline{Y}$ , $\overline{Z}$ , sobre el mismo plano, porque $\overline{Y}$ es la imagen de $\overline{V}$ de algún múltiplo escalar de $\overline{Z}$ . Así $\overline{Z}$ es una combinación lineal de $\overline{V}$ y. $\overline{Y}$ . En consecuencia $\overline{X}\cdot\overline{V} = 0$ y $\overline{Y}\cdot\overline{X} = 0$ , entonces $\overline{X}\cdot\overline{Z} = 0$ . La normalización de cualquier vector $\overline{X}'$ satisface las ecuaciones $\overline{X}' \cdot \overline{V} = 0$ y $\overline{X}' \cdot \overline{Z} = 0$ será un vector unitario sobre el eje $x'$ en *P**, $\overline{X}' = \overline{(-v_2, v_1, 0)}$ satisface estas dos ecuaciones. Sea $d_1 = \sqrt{v_1^2 + v_2^2}$ entonces

$$\overline{X}' = \overline{(-v_2 / d_1, v_1 / d_1, 0)}$$

es el vector unitario positivo sobre el eje $x'$ de *P**. Las coordenadas de $\overline{Y}$ son determinadas similarmente de la ecuación $\overline{Y} \cdot \overline{X} = 0$ y $\overline{Y} \cdot \overline{V} = 0$. Sea $d = \sqrt{v_1^2 + v_2^2 + v_3^2}$ . Entonces

$$Y = \overline{\left(-v_1 v_3, -v_2 v_3, d_1^2\right) / d d_1}$$

es el vector unitario positivo sobre el *y'* de *P**. Sustituyendo estos valores de $\overline{X}$ y $\overline{Y}$ en las ecuaciones (4) y (5), tenemos

$$u = r\left(a_2 v_1 - a_1 v_2\right) / d_1 \tag{7}$$

$$v = r\left(-a_1 v_1 v_3 - a_2 v_2 v_3 + a_3 d_1^2\right) / \left(d d_1\right) \tag{8}$$

donde, $r = d^2 / \left(d^2 - \left(a_1 v_1 - a_2 v_2 + a_3 v_3\right)\right)$ como definió en ecuación. (6).

La expresión anterior para $v$ es algebraicamente equivalente lo siguiente, que involucra algunas operaciones aritméticas:

$$v = \left(v_3 + r\left(a_3 - v_3\right)\right) \cdot d / d_1$$

Así, dado un punto $A$ sobre *S*, las coordenadas rectangulares *(u,v)* de este punto en la imagen son la evaluadas por las ecuaciones (7) y (9).

REFERENCIAS